*Communication*

# A Simplistic and Cost-Effective Design for Real-World Development of an Ambient Assisted Living System for Fall Detection and Indoor Localization: Proof of Concept

**Nirmalya Thakur [1]* and Chia Y. Han [2]**

[1] Department of Electrical Engineering and Computer Science, University of Cincinnati, Cincinnati, OH 45221-0030, USA; thakurna@mail.uc.edu
[2] Department of Electrical Engineering and Computer Science, University of Cincinnati, Cincinnati, OH 45221-0030, USA; han@ucmail.uc.edu
* Correspondence: thakurna@mail.uc.edu

**Abstract:** Falls, highly common in the constantly increasing global aging population, can have a variety of negative effects on their health, well-being, and quality of life, including restricting their capabilities to conduct Activities of Daily Living (ADLs), which are crucial for one's sustenance. Timely assistance during falls is highly necessary, which involves tracking the indoor location of the elderly during their diverse navigational patterns associated with ADLs to detect the precise location of a fall. With the decreasing caregiver population on a global scale, it is important that the future of intelligent living environments can detect falls during ADL.s while being able to track the indoor location of the elderly in the real world. Prior works in these fields have several limitations, such as – the lack of functionalities to detect both falls and indoor locations, high cost of implementation, complicated design, the requirement of multiple hardware components for deployment, and the necessity to develop new hardware for implementation, which make the wide-scale deployment of such technologies challenging. To address these challenges, this work proposes a cost-effective and simplistic design paradigm for an Ambient Assisted Living system that can capture multimodal components of user behaviors during ADLs that are necessary for performing fall detection and indoor localization in a simultaneous manner in the real world. Proof of concept results from real-world experiments are presented to uphold the effective working of the system. The findings from two comparison studies with prior works in this field are also presented to uphold the novelty of this work. The first comparison study shows how the proposed system outperforms prior works in the areas of indoor localization and fall detection in terms of the effectiveness of its software design and hardware design. The second comparison study shows that the cost for the development of this system is the least as compared to prior works in these fields, which involved real-world development of the underlining systems, thereby upholding its cost-effective nature.

## 1. Introduction

Growing worldwide longevity is now more commonplace than ever before, with the average life expectancy reaching 60 years or higher. This is mostly due to medical breakthroughs and advances in healthcare research [1]. The population of the world over the age of 65 is increasing dramatically, numbering 962 million today, and projected to increase to 2 billion by 2050 [2,3]. As the population ages, modern society is facing a wide range of difficulties stemming from numerous conditions associated with the elderly, such as varying rates of decline in behavioral, social, emotional, mental, psychological, and motor abilities, as well as other issues such as cognitive impairment, behavioral disorders, disabilities, neurological disorders, Dementia, Alzheimer's, and visual impairments, that are associated with the process of aging.

Over the last few years, the aging populations throughout the globe have had to contend with a decrease of caregivers to care for them, which has created a variety of challenges and difficulties [5-7]. Here, two of the key challenges will be discussed. First, as the demand has increased, the cost of caregiving has risen considerably in recent years. As a result, affording caregivers is becoming increasingly difficult. Second, quite often, caregivers take care of multiple elderly people with multiple varying needs during a day; as a result, they are frequently exhausted, overworked, overwhelmed, and overburdened, which affects the quality of care.

Research predicts that the worldwide population comprising both the elderly and the young will live in smart homes, smart communities, and smart cities in the coming years. Research by [8] estimates that 66 percent of the world's population will live in smart homes by 2050. Thus, due to the scarcity of caregivers and the expected emergence of smart homes on a global scale [9], the future of technology-driven Internet of Things (IoT)-based living spaces, such as smart homes, must be able to contribute to Ambient Assisted Living (AAL) for the elderly by detecting, interpreting, analyzing, and anticipating different needs within the context of their ADLs.

AAL may broadly be defined as the use of networked, automated, and/or semi-automated technological solutions within people's living and working surroundings to improve their health and well-being, quality of life, user experience, and independence [10]. In general, ADLs may be considered as the everyday activities needed for one's sustenance done in one's living surroundings [11]. Categories of ADLs include personal hygiene, dressing, eating, continence management, and mobility.

As one becomes older, one becomes more prone to falling. As per [12], a fall is defined as a sudden drop onto the ground or floor resulting from being pushed or pulled, environmental factors, fainting, or any other analogous health-related issues, difficulties, or impairments. Falls can have many consequences on the health, well-being, and quality of life of the elderly, such as making it difficult for them to complete ADLs. On a worldwide scale, falls are the second most common cause of unintentional fatalities. Older individuals are at an increased risk of suffering a traumatic brain injury due to falling [13]. On an annual basis, around one in every three older persons fall at least once a year, and it is estimated that the percentage of those who fall will rise by around 50% soon [14,15]. Falls have been a major concern for the worldwide elderly population. In the United States alone, since 2009, the number of people who have died from falls has increased by 30%. In the United States, every 11 seconds, an elderly person who has fallen has to be sent to the hospital for urgent care; an older adult dies every 19 minutes from a fall; every year, there are approximately 3 million emergency room visits, 800,000 hospitalizations, and more than 32,000 fatalities among the elderly due to falls. The rate of injuries and fatalities from falls is rising consistently. By 2030, in the United States, research predicts that there will be seven deaths per hour from falls. The yearly cost of medical and healthcare-related costs connected to falls among the elderly is USD 50 billion. This figure is expected to grow not only in the US but also on a global scale [16-18]. A fall can be caused by various reasons, which can be roughly classified as internal and external. External causes are related to the environmental variables in the spatial confines of the individual that might contribute to a fall. These include slippery surfaces, staircases, and so forth. Causes such as impaired eyesight, cramping, weakening in muscular skeleton structure, chronic diseases, and so on are examples of internal reasons for a fall. In addition to causing mild to serious physical injuries, falls can have a wide range of negative effects on the individual, which may include individual—injuries, bruising, blood clots; social life—reduced mobility leading to loneliness and social isolation; cognitive or mental—fear of moving about, lack of confidence in doing ADLs; and financial—the expenses of medical treatment and caretakers [19].

To accurately detect falls and the other dynamic and diversified needs of the elderly that usually arise in the context of their living environments during ADLs, tracking and analysis of the indoor spatial and contextual data associated with these activities are highly crucial. Technologies such as Global Positioning Systems (GPS) and Global Navigation Satellite Systems (GNSS) have transformed navigation research by allowing people, objects, and assets to be tracked in real time. Despite their great success in outdoor contexts, these technologies are still unsuccessful in indoor settings [20]. This is due to two factors: first, these technologies rely on line-of-sight communication between GPS satellites and receivers, which is not achievable in an indoor setting, and second, GPS has a maximum accuracy of up to five meters [21]. An Indoor Localization System is a network of systems, devices, and services that assist in tracking and locating persons, objects, and assets in indoor environments where satellite navigation systems like GPS and GNSS are ineffective [22]. Thus, Indoor Localization becomes highly relevant for AAL so that the future of technology-based living environments such as Smart Homes can take a comprehensive approach towards addressing the multimodal and diverse needs of the elderly during different ADLs as and when such needs arise.

Despite recent advances in AAL research, when considering the development of smart home technologies, numerous challenges remain, as mentioned in detail in Section 2. These challenges are primarily centered around - the lack of real-world testing, lack of functionalities in the frameworks to detect both falls and indoor locations in a simultaneous manner, high cost of implementation, complicated design paradigms, the requirement of multiple hardware components for deployment, and the necessity to develop new hardware or software for implementation, which make the wide-scale deployment of such technologies challenging. With an aim to address these challenges, our paper makes the following scientific contributions to this field:

1. It presents a simplistic design paradigm for an AAL system that can capture multimodal components of user behaviors during ADLs that are necessary for performing fall detection and indoor localization in a simultaneous manner in the real world. A comprehensive comparative study (Table 3) with prior works in the fields of indoor localization and fall detection is presented in this paper, which shows how this proposed system outperforms prior works in the fields of indoor localization and fall detection in terms of the effectiveness of software design and hardware design.
2. The development of this system is highly cost-effective. We present a second comparative study (Table 4) where we compare the cost of our system with the cost of prior works in these fields, which involved real-time development. This comparative study upholds the fact that the cost of our system is the least as compared to all these works, thereby upholding its cost-effective nature. For this comparative study, we used only the cost of equipment as the grounds for comparison. While there can be several other costs (such as cost of installation, cost of maintenance, the salary of research personnel, cost of deployment, computational costs, etc.) that can be computed, but most of the prior works in this field reported only the cost of equipment, so only this parameter was used as the grounds for comparison in this comparative study. Furthermore, comparing the cost of the associated equipment to comment on the cost-effectiveness of the underlying system is an approach that has been followed by researchers in the broad domain of IoT as can be seen from recent works in this field (Section 4.2).

The rest of this paper is presented as follows. Section 2 reviews the recent works in this field and elaborates on the challenges and limitations that exist in these systems. Section 3 presents the methodology and concept for the development of the proposed system. Section 4 presents the results and discussions from the real-world testing of the proposed system. It also includes the above-mentioned comparison studies. It is followed by Section 5, which concludes the paper and outlines the scope for future work in this field.

## 2. Literature Review

In this section, a comprehensive review of recent works in Indoor Localization and Fall Detection that focused on Ambient Assisted Living is presented.

Varma et al. [23] developed an indoor localization system using a Random Forest-driven machine learning approach based on data gathered from 13 beacons installed in a simulated Internet of Things (IoT) infrastructure. To get the user's location, the researchers processed the data from all these beacons. The development of a neural network-based indoor WiFi fingerprinting method by Qin et al. [24] enabled a user's position to be pinpointed in an indoor setting. They examined their findings on two datasets to discuss the effectiveness of their work. The user's position was detected using a decision tree-based technique in the research proposed by Musa et al. [25]. The system design included a non-line-of-sight approach, multipath propagation tracking, and ultra-wideband technique. Similarly, a decision tree-based method for indoor location detection was presented by Yim et al. [26]. The system was able to use WiFi fingerprinting data to build the decision tree offline. According to Hu et al. [27] 's technique, a k-NN classification methodology with contextual data from an indoor environment can locate a person's indoor position by using the access point the person was closest to after accessing it. Poulose et al. [28] developed a deep learning-based method for indoor position tracking that employed Received Signal Strength Indicator (RSSI) signal data to train the learning model. Barsocchi et al. [29] utilized a linear regression-based learning technique to build an indoor positioning system that used RSSI values to track the user's distance from reference locations and subsequently translated the same numerical value to a distance measure to locate the user's real position. Kothari et al. [30] created a cost-effective, user-location-detecting smartphone application. Four volunteers were included in the trials to evaluate the technique, which merged dead reckoning with WiFi fingerprinting. Wu et al. [31] created an indoor positioning system using data from new sensors incorporated into the users' mobile phones that could utilize user motion and user behavioral features to create a radio map with a floor plan, which could subsequently be used to determine the users' indoor positions. The researchers recruited a total of 4 participants to evaluate their framework. Gu et al. [32] suggested a step counting method that could address challenges such as over-counting steps and false walking while tracking the user's indoor location; that was validated by taking into consideration the data collected from 8 participants. Similar design paradigms were followed by researchers in [33-35] to test the underlining systems. In addition to these approaches for indoor localization, the concept of optical positioning has also been explored in some research works [36-42] in this field. The success of the optical positioning approach is centered around the use of optical positioning sensors to detect the indoor locations of people and objects.

Following the above, we examine recent AAL works related to fall detection. An algorithm designed by Rafferty et al. [43] used thermal vision to track falls during ADLs. The system architecture involved installing thermal vision sensors on the ceiling in the confines of the user's living space, and then computer vision algorithms were used to identify falls. In the work of Ozcan et al. [44], the camera had to be carried by the user instead of being installed at various locations. A decision tree-based technique was utilized to identify falls based on images taken by the user's camera. A wearable device for fall detection was developed by Khan et al. [45]. This gadget had a three-part assembly: a camera, a gyroscope, and an accelerometer, all of which were connected to a computer that comprised the system architecture to detect falls. The work of Cahoolessur et al. in [46] introduced a binary classifier-based device capable of finding anomalies in behavioral patterns such as falls in a simulated IoT-based environment. To design the wearable gadget, the authors first developed a model for the user, using a cloud computing-based architecture, which was followed by implementation and testing of the same. Godfrey et al. [47] unified algorithms for fall detection and gait segmentation for developing a framework to assess the free living of individuals suffering from Parkinson's disease. Liu et al. [48] used Doppler sensors to develop a fall detection system with a specific focus on detecting falls in senior citizens. In [49] and [50], Dinh et al. proposed a novel wearable device and a system architecture that used inertial sensors and Zigbee transceivers to detect falls. The works in [51] and [52] proposed somewhat similar system paradigms as compared to the previous works to track the positions and health status of individuals for their assisted living. Hsu et al. [53] developed a backpropagation neural network that used the Gaussian Mixture Model (G.M.M.) to detect falls during various forms of movement patterns. In [54], Yun et al. developed a smart camera-based system for fall detection. The system worked by detecting falls from a single camera with arbitrary view angles. Nguyen et al. [55] proposed a computer vision-based system for fall detection that was characterized by low power consumption during its operation. In [56], Huang used a combination of ultrasonic sensors and field-programmable gate array (FPGA) processors to develop a system architecture to monitor the postures and motions of individuals to detect falls. In [57], the authors proposed UbiCare, which was a system developed by integration of several commercially available sensors such as Arduino microcontrollers and ZigBee communicators for the assisted living of the elderly during ADLs.

In addition to these works on indoor localization and fall detection, there has been a significant amount of work in these fields where only datasets were used to test and validate the proposed software designs and/or software frameworks. These include the works in indoor localization by Song et al. [58], Kim et al. [59], Jang et al. [60], Wang et al. [61], Qin et al. [62], Wietrzykowski et al. [63], Panja et al. [64], Yin et al. [65], Patil et al. [66], Gan et al. [67], Hoang et al. [68], and Seçkin et al. [69]. In the field of fall detection, such works include the systems proposed by – Galvão et al. [70], Sase et al. [71], Li et al. [72], Theodoridis et al. [73], Abobakr et al. [74], Abdo et al. [75], Sowmyayani et al. [76], Kalita et al. [77], Soni et al. [78], Serpa et al. [79], and Lin et al. [80].

As can be seen from these recent works [23-80] in the field of AAL, with a specific focus on fall detection and indoor localization, the following challenges exist in these systems:

1. The available AAL-based systems for fall detection cannot track the user's indoor location and vice versa. Furthermore, the hardware components (sensors) that were used to develop the fall detection systems cannot be programmed or customized to capture the necessary data required for incorporating the functionality of indoor localization in such systems and vice versa. For instance, a host of beacons [23], WiFi access points [26], and WiFi fingerprint capturing architecture [27] help to capture the necessary data for indoor localization but these hardware components cannot be programmed or customized to capture any relevant data that would be necessary for detecting falls. It is highly essential that, in addition to being able to track, analyze, and interpret human behavior, such systems are also able to detect the associated indoor location so that the same can be communicated to caregivers or emergency responders to facilitate timely care in the event of a fall or any similar health-related emergencies. Delay in care from a health-related emergency, such as a fall, can have both short-term and long-term health-related impacts.
2. A majority of these systems were tested on datasets [24, 58-80]. The proposed software designs and/or software frameworks that were used by the authors of these respective works cannot be directly applied in the real world to detect falls and indoor locations of users during ADLs performed in real-time as the underlining systems were not developed to work based on incoming or continuously generating human behavior-based data from the real-world.
3. While there have been some works that have involved the real-world implementation of the underlining AAL systems, the cost of equipment necessary for the development of such systems is very high. For instance, the systems proposed by: Kohoutek et al. [36] costs USD 9000, Muffert et al. [37] costs more than USD 10000, Tilsch et al. [38] costs about USD 1050, Habbecke et al. costs about USD 1050, Popescu et al. [40] costs USD 1500, Huang et al. [56] costs USD 750, Dasios et al. [57] costs USD 581, and so on. Such high costs are a major challenge to the real-world development and wide-scale deployment of such systems across multiple smart homes.

4. These methodologies use multiple sensors and hardware systems that need to be installed in the living confines of the user. Some examples include - 13 beacons [23], WiFi access points and WiFi fingerprint capturing architecture [26,27], RSSI data capturing methodologies [28,29], thermal vision sensors [43], and smart cameras [44] that need to be carried by the users. Installing such sensors across smart communities or smart cities that could represent multiple interconnected smart homes would be highly costly, and the elderly are usually receptive to the introduction of such a host of hardware components in their living environments [81].
5. The design process for the development of most of these systems [23,26-29,33-42,47-57] is complicated as it involves the integration and communication of multiple software and hardware components. As there is a need for the development of AAL systems that can perform both fall detections and indoor localizations in a simultaneous manner in the real world, so integrating the hardware components of these underlining systems (integrating hardware components from systems aimed at fall detection with hardware components from systems aimed at indoor localization) and developing a software framework that can communicate, share, and exchange data with all these hardware components in a seamless manner in real-time would be even more complicated.
6. Some of the works have also involved the development of new applications – such as the smartphone-based application proposed in [30] and the wearable devices proposed in [46], [49], and [50]. Replicating the design of an application has several challenges unless it is replicated or re-developed by the original developers [82]. In the context of wearables, it is crucial to ensure that the design methodology follows the 'wearables for all' design approaches [83], which poses a challenge to the mass development of such wearable solutions.

To address the above limitations, we propose a cost-effective and simplistic design paradigm for an Ambient Assisted Living system that can capture multimodal components of the user behavior during ADLs that are necessary for performing fall detection and indoor localization in a simultaneous manner in the real world. The system design and the associated methodology are discussed in the next section.

## 3. Methodology and System Design

This section outlines the design process for the development of the proposed AAL system. The design process is based on the findings of two of our recent works [84,85] in these fields. These works [84,85] were centered around the development of systems for detecting falls and tracking the indoor location of users during ADLs. Various forms of user interaction data, user behavior data, and user posture data were studied in these works. The proposed system architectures were tested on different datasets, and the findings from both these works prove that the accelerometer and gyroscope data (in X, Y, and Z directions) collected from wearables during different ADLs can be studied, analyzed, interpreted, and utilized to develop a machine learning-based classifier that can detect falls as well as track indoor locations of users with a very high level of accuracy.

In this paper, we extend both these works to develop a cost-effective and simplistic design paradigm for an Ambient Assisted Living system that can capture multimodal components of user behaviors during ADLs that are necessary for performing fall detection and indoor localization in a simultaneous manner in the real world. These respective functionalities for fall detection and indoor localization are outlined in 3.1 and 3.2, respectively. The proposed design and the associated system specifications that integrate both these functionalities [84,85] as a software solution for a real-world environment are presented in Section 3.3.

### 3.1. Methodology for Fall Detection during ADLs

This approach's design specification [84] comprises four main components: pose detection, data collecting and preprocessing, learning module, and performance module. To identify the user's position at any given time, the pose detection system utilizes motion and movement-related data during ADLs to deduce the user's instantaneous position. The study of accelerometer data during various activities at various timestamps is one way of creating such a pose identification system. This is measured by computing the acceleration vector and the acceleration's orientation angles measured on all three axes. Every posture, motion, and movement may be investigated by computing these orientation angles as a unique spatial orientation during different stages or steps of any given ADL. In this case, the location of the sensor on the body is equally essential. An accelerometer positioned on the chest of a person, for instance, is always at 90 degrees with the floor for motions like standing and sitting but at 0 degrees with the floor for other types of movements like lying on the ground or being in a stance in which both arms and legs are touching the floor. In analyzing postures or poses related to various ADLs, the purpose is to compute orientation angles along the three axes. If the user is moving, they will adopt dynamic postures. The rapid changes in the values of the acceleration angles can be used to identify falls or motions associated with falls. The approach for collecting and preparing data is part of the data collection and preprocessing module of this system. The learning module of the system comprised the AdaBoost approach with the cross-validation operator applied to a kNN classifier. The performance module evaluated the performance of the learning module using a confusion matrix. It was tested on two datasets [87,88]. The system achieved performance accuracies of 99.87% and 99.66%, respectively, when tested on these two datasets. The study did not report any hardware system design for real-world development. Neither did it report any real-world data collection methodology that would be necessary for such a system to function in real-time. The findings also showed that the accelerometer data and gyroscope data (from X, Y, and Z directions) comprise the necessary attributes that are needed to develop such a system that can outperform all prior works in the field of fall detection in terms of performance accuracy (the detailed comparison study with prior works in fall detection has been presented in [84]).

### 3.2. Methodology for Indoor Localization during ADLs

The development of the proposed approach [85] for acceleration and gyroscope data-based indoor localization during ADLs involves the following steps:

1. The associated representation scheme involves mapping the entire spatial location into non-overlapping 'activity-based zones' [85], distinct to different complex activities, by performing complex activity analysis [89].
2. The complex activity analysis as per [89] involves detecting and analyzing the ADLs in terms of the atomic activities, context attributes, core atomic activities, core context attributes, start atomic activities, start context attributes, end atomic activities, and end context attributes using probabilistic reasoning principles and the associated weights of each of these components of a given ADL.
3. Inferring the semantic relationships between the changing dynamics of these actions and the context-based parameters of these actions.
4. Studying and analyzing the semantic relationships between the accelerometer data, gyroscope data, and the associated actions and the context-based parameters of these actions (obtained from the complex activity analysis) within each 'activity-based zone'.
5. Interpreting the semantic relationships between the accelerometer data, gyroscope data, and the associated actions and the context-based parameters of these actions across different 'activity-based zones' based on the sequence in which the different ADLs took place and the related temporal information.

6. Integrating the findings from Step 4 and Step 5 to interpret the interrelated and semantic relationships between the accelerometer data and the gyroscope data with the location information associated with different ADLs that were successfully completed in all the 'activity-based zones' in the given indoor environment.
7. Splitting the data into the training set and test set and developing a machine learning-based model to detect the location of a user in terms of these spatial 'zones' based on the associated accelerometer data and gyroscope data.
8. Computing the accuracy of the system by using a confusion matrix.

Based on the above steps, the system was developed by using a Random Forest classifier. The effectiveness of the system was tested using a dataset [88]. The study did not report any hardware system design for real-world development. Neither did it report any real-world data collection methodology that would be necessary for such a system to function in real-time. The system achieved an overall performance accuracy of 81.13% for detecting the indoor location of a user in different activity-based zones such as bedroom zone, kitchen zone, office zone, and toilet zone. The findings also showed that the accelerometer data and gyroscope data (from X, Y, and Z directions) during ADLs can be used to develop a context-independent indoor localization system that can work effectively to detect the indoor location of a user in different IoT-based settings.

### 3.3. Methodology for Indoor Localization and Fall Detection: Real-World Implementation

The development of the proposed system, including its design and specifications by incorporating the functionalities for fall detection [84] and indoor localization [85] (outlined in Sections 3.1 and 3.2, respectively), is presented in this section. The system, proposed as a software solution, functioned by following a design paradigm that involved successful integration, communication, and interfacing between multiple sensors that were characterized by their capabilities to capture the necessary data (as outlined in Sections 3.1 and 3.2, respectively). Upon obtaining this necessary data, the system would be able to extract the interdependent and multi-level semantic relationships between user interactions, context parameters, and daily activities concerning the dynamic spatial, temporal, and global orientations of a user during different ADLs. This system was developed, implemented, deployed, and tested at the Ambient Assisted Living Research Lab located at 411 Science Building, University of Cincinnati Victory Parkway Campus.

The design paradigm of the proposed system involved following a simplistic and cost-effective approach towards integration, communication, and interfacing between multiple off-the-shelf sensors. These off-the-shelf sensors comprised the following:

(a) The Imou Bullet 2S Smart Camera: The Imou Bullet 2S is a smart camera that can directly connect to WiFi and can be used to capture different components of the video-based and image-based data during different ADLs. It has features such as infrared mode, color mode, smart mode, and human detection. The technical specifications of this smart camera include 1080P Full HD glass optics, 2.8 mm lens, 120° viewing angle, 98ft night vision, IR lighting, inbuilt image-processing algorithm, storage facility via the H.265 compression system on an SD card (up to 256 GB) or on an encrypted cloud server, human motion detection, and inbuilt microphone [90].

(b) The Sleeve Sensor Research Kit: The Sleeve Sensor Research Kit has several components to record the different characteristics of motion and behavior data during ADLs. These include an Accelerometer, Gyroscope, Magnetometer, Sensor Fusion, Pressure Sensor, and Temperature Sensor. Specifically, this MMS sensing system consists of a wearable device with the following sensors: 6-axis Accelerometer + Gyroscope, BMI270 Temperature, BMP280 LTR-329ALS, BMP280 Barometer/Pressure/Altimeter, Ambient Light/Luminosity Magnetometer, with three axes, BMM150 Sensor Fusion, 9-axis BOSCH 512MB memory, Lithium-ion rechargeable battery, Bluetooth Low Energy, CPU, button, LED, and GPIOs [91].

(c) The Estimote Proximity Beacons: The Estimote Proximity Beacons can be used to track the proximity of a user to different context parameters as well as to detect the presence or absence of the user in a specific 'activity-based zone' during different ADLs. Each beacon has a low-power ARM® CPU (e.g., 32-bit or 64 MHz CPU), a quad-core, 64-bit, 1.2 GHz CPU in Mirror flash memory to store apps and data, 8 GB in Mirror RAM memory for the apps to use while running, 1 GB in Mirror, a Bluetooth antenna and chip to communicate with other devices, and between the beacons themselves [92].

Based on the system paradigms described in Sections 3.1 and 3.2 and the physical dimensions of the Ambient Assisted Living lab, 2 Imou Bullet 2S Smart Cameras, 1 Sleeve Sensor Research Kit, and 4 Estimote Proximity Beacons were used for developing the proposed system. In addition to these sensors, Microsoft SQL Server version 11.0 [93] was used to develop the database, which is an integral part of the proposed software solution. Microsoft SQL Server is a Relational Database Management System (RDBMS) developed by Microsoft. These sensors were specifically chosen for the following reasons:

1. These sensors can be programmed to capture the specific components of user behavior and user posture data (as evidenced in our prior works – [84] and [85]) that are necessary to detect both falls and indoor locations during ADLs.
2. They are cost-effective.
3. These sensors are easily available and can be seamlessly set up in any given indoor space or region without any training.
4. The development of a software solution that can communicate and interface with all these sensors is not complicated.
5. The design process, both for the experiments and for the system architecture, becomes convenient due to the specifications, coverage, and characteristics of these sensors.

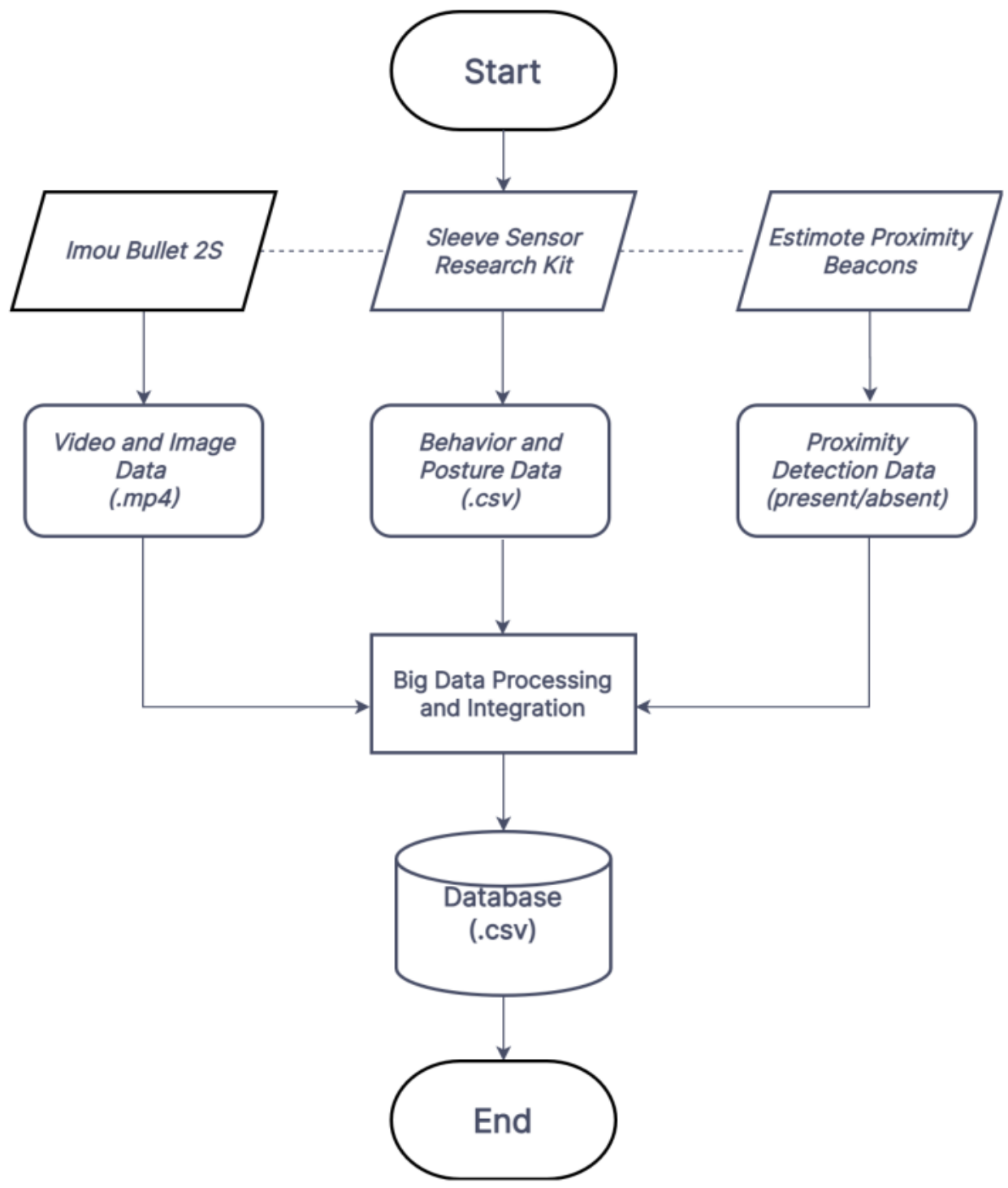

**Figure 1** Flowchart illustrating the working of the proposed system

The simplistic nature of this design process as compared to all prior works in this field is explained next. As can be seen from Section 2, the works that involved real-world fall detection or indoor localization involved multiple sensors and/or hardware equipment. For instance – as per recent works in these fields, some of the most commonly used sensors and/or hardware equipment for indoor localization include - 13 beacons (Varma et al. [23]), RSSI transmitters and receivers (Poulose et al. [28] and Barsocchi et al. [29]), system setup for capturing WiFi fingerprints (Yim et al. [26] and Kothari et al. [30]) as well as the development of completely new sensors (Wu et al. [31]). Similarly, some of the most commonly used sensors and/or hardware equipment for fall detection include - thermal vision cameras (Rafferty et al. [43]), smart cameras (Ozcan et al. [44] and Yun et al. [54]), Doppler sensors (Liu et al. [48]), Inertial Sensors and Zigbee transceivers (Dinh et al. [49] and [50]), and ultrasonic sensors and FPGA processors (Huang [56]). The sensors that were used for indoor localization cannot be programmed or customized to detect falls. Similarly, the sensors that were used to detect falls cannot be programmed or customized to track the indoor locations of users. We discussed the need for the future of AAL systems to be able to detect both falls and indoor locations in a simultaneous manner in real-time (Section 1). Therefore, based on these recent works, such an AAL system would involve the integration of sensors used in any of the works for indoor localization (for instance – [23], [28], or [29]) with the sensors used in any of these works for fall detection (for instance – [43], [44], [54], [48], [49], [50], [56]). An example of such a system could include a system architecture that comprises a host of 13 beacons for indoor localization (if the indoor localization methodology proposed in [23] is used) and a host of inertial sensors and Zigbee transceivers mounted throughout the living space of an individual (if the fall detection methodology proposed in [49] or [50] is used). Thereafter, the software design also has to be customized so that the underlining software can seamlessly communicate and process the real-time data coming from these host of hardware components or sensors. Such a system setup (both in terms of integrating the hardware components and as well as for setting up the software component to seamlessly communicate with all the hardware components) is going to be much more complicated as compared to the integration of a few off-the-shelf sensors (2 Imou Bullet 2S Smart Cameras, 1 Sleeve Sensor Research Kit, and 4 Estimote Proximity Beacons) that can be programmed to capture the required components of user behavior and user posture data that are necessary to detect falls as well as detect indoor locations during ADLs in a simultaneous manner in the real world. This helps to uphold the simplistic design paradigm of the proposed system. This is further explained in the rest of this section.

The flowchart for the development of our system is shown in Figure 1. As can be seen from Figure 1, the system comprises of three main components – data capture, data processing and integration, and database development. The data capture process was centered around activation, calibration, and utilization of the sensor components to start capturing the data. This process comprised of the video data collected by the Imou Bullet 2S Smart Camera's in .mp4 format, the accelerometer and gyroscope data collected by the Sleeve Sensor Research Kit available as different .csv files, and the proximity detection data in terms of presence or absence of the user in a specific activity-based zone collected by the Estimote Proximity Beacons. As there were four proximity beacons that were set up, so, at a given point of time when the user was present in a specific activity-based zone, the proximity beacon of that zone would mark the user as present, and the other three proximity beacons (present in three other activity-based zones) would mark the user as absent in those respective zones. The next module of this system was the big data processing and integration module. The functionality of this module was characterized by preprocessing and integration of the data from the different sensors in one format. This module also involved the capability to successfully communicate with different sensors while each of these sensors was collecting data in real-time. This module was developed as an application using the C# programming language in the .NET framework on a Windows 10 computer with an Intel (R) Core (TM) i7-7600U CPU @ 2.80 GHz, two core(s), and four logical processor(s). While other programming languages could also have been used, C# was used in view of the developer features provided by these respective sensor platforms, as well as the ease of integration and communication between applications developed on C# and .NET and Microsoft SQL Server 2012, version 11.0. The next module, database development, was centered around the development of a database on Microsoft SQL Server 2012, version 11.0 [93]. Microsoft SQL Server 2012, version 11.0, is available for free download at the link mentioned in [93], so no additional expenses were incurred for setting up this database on a local server. The database comprised different attributes that represented the different characteristics of the data obtained from the different sensor components upon the successful integration of the same. As this is a proof of concept study, here we are demonstrating the ability of our proposed system to capture the specific components of user interaction, user behavior, and user posture data (as evidenced in prior works – [84] and [85]) that are necessary for performing fall detection and indoor localization during ADLs in a real-world scenario. In the near future, we will be adding another module to the system, which would be the detection module. This module would comprise the specific systems for fall detection and indoor localization (as proposed in prior works [84] and [85]) to enable those systems to function on real-time data.

Figure 2 is a screenshot from the user interface of the software, which shows the real-time data collection from the Sleeve Sensor Research Kit from Mbient Labs during one of the experiments. Figure 3 shows the Sleeve Sensor Research Kit. This image is provided separately as due to the small size of the sensor, it is difficult to track the location of the sensor in the images from the experiments, which are provided in Section 4. During the experiments, as per the methodology described in Sections 3.1 and 3.2, this sensor was mounted on the user's chest to collect the data during different ADLs performed in different 'activity-based zones'. Figures 4 and 5 show the placement of multiple proximity sensors in different 'activity-based zones'. Figures 6 and 7 show the strategic and calculated placement of the two Imou smart cameras, which helped to map the entire available space in the research lab.

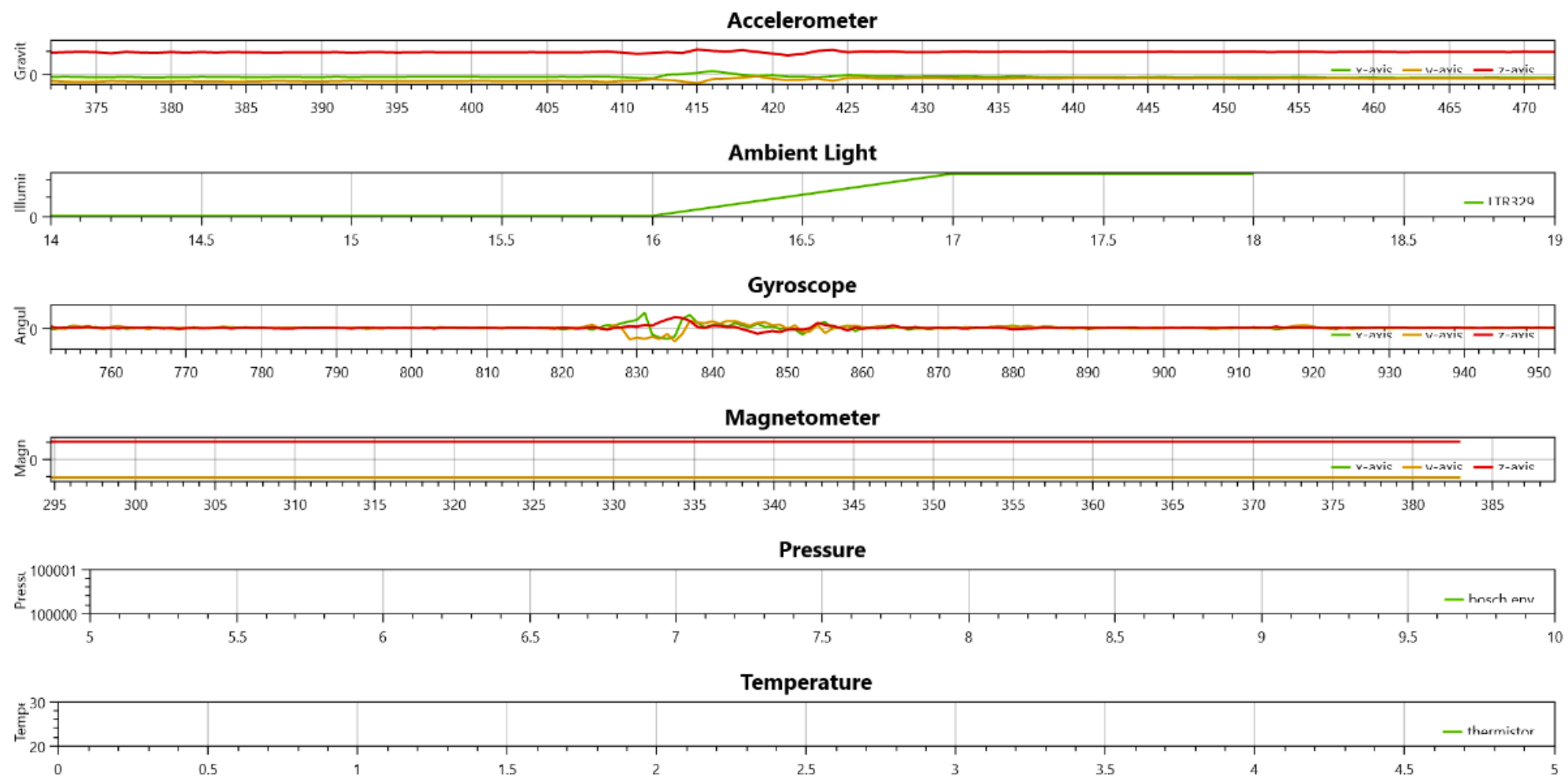

**Figure 2**. Real-time data collection from the Sleeve Sensor Research Kit from Mbient Labs during one of the experiments

As can be seen from Figures 6 and 7, dedicated strategic locations (marked as Camera 1 zone and Camera 2 zone) were assigned in the lab space for the setting up of both these Imou Smart cameras. The cost of all these sensors (at the time of purchase) is outlined in Table 1. In Section 4, we compare this cost to the cost of similar fall detection, indoor localization, and assisted living solutions that have been proposed in prior works in this field to justify the cost-effectiveness of this proposed approach.

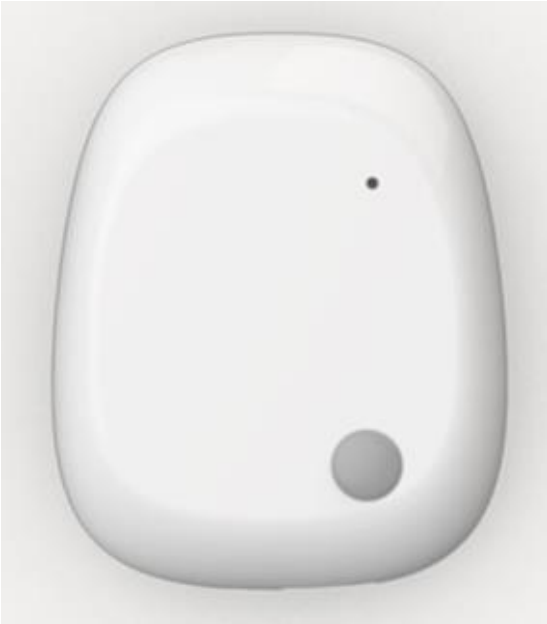

**Figure 3**. The Sleeve Sensor Research Kit

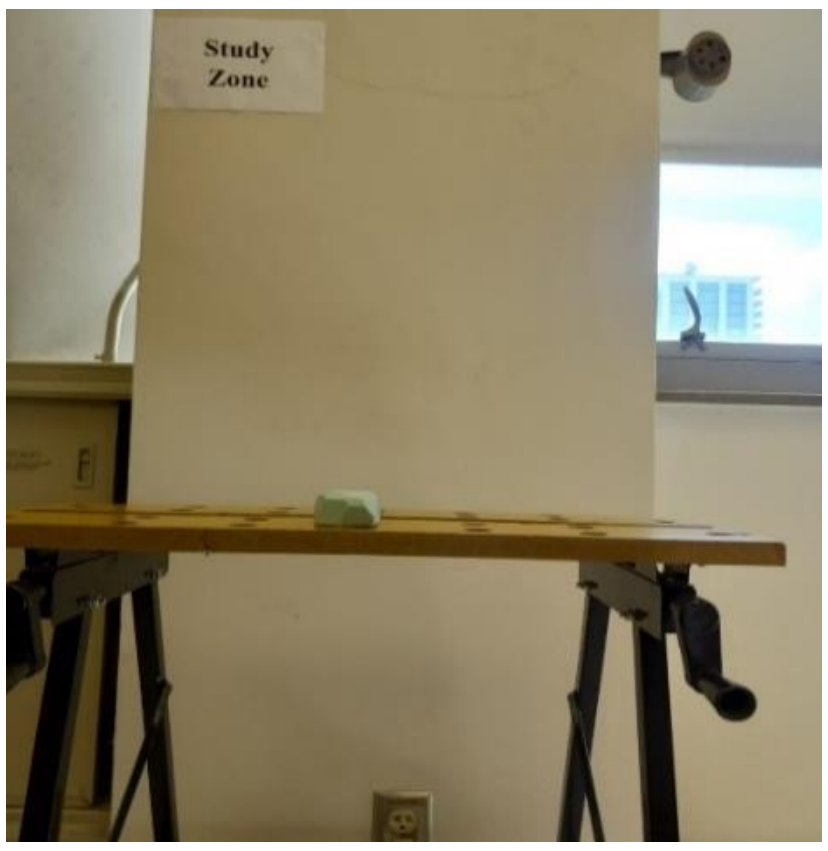

**Figure 4**. Placement of one of the proximity sensors in the 'Study-Zone.

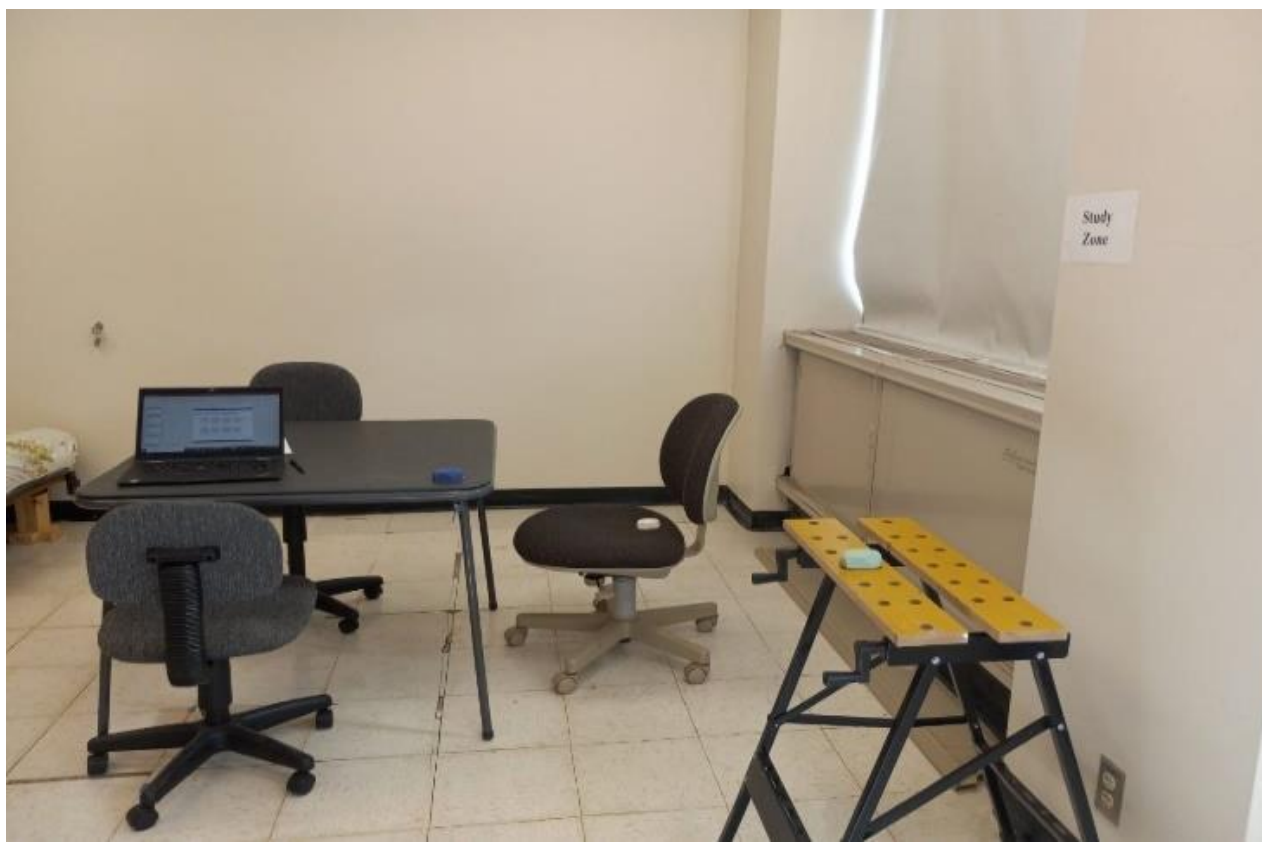

**Figure 5**. Placement of multiple proximity sensors in the 'Study-Zone.'

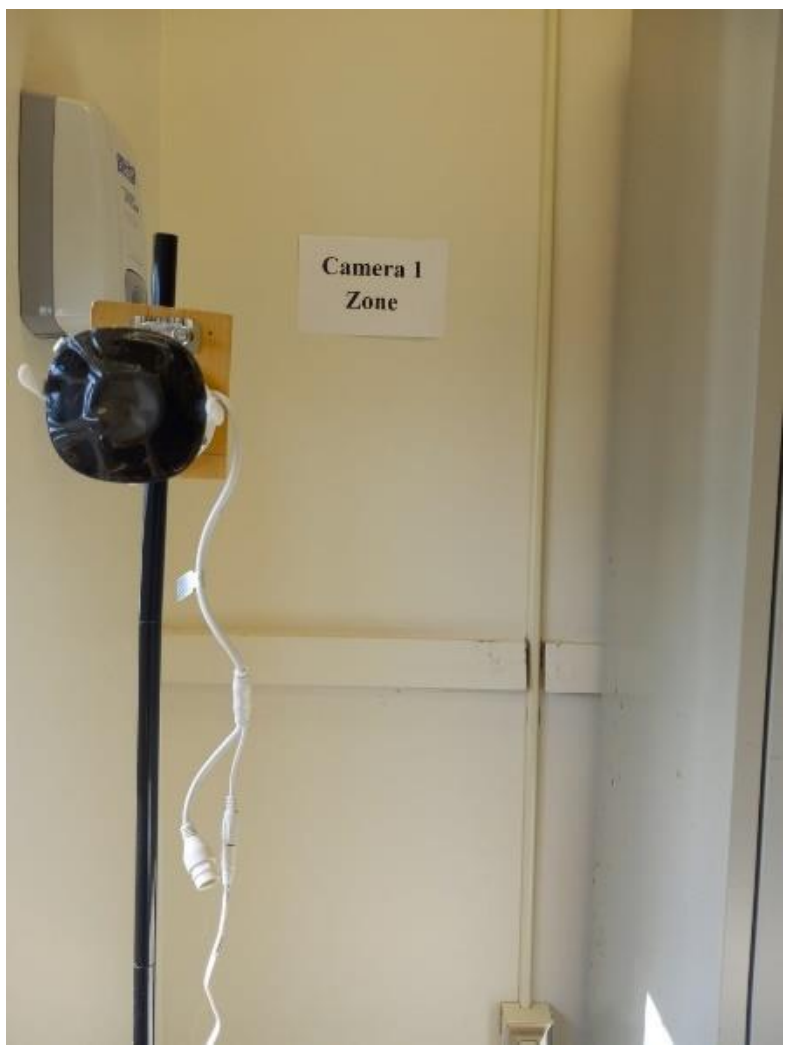

**Figure 6**. Placement of one of the Imou Smart Cameras

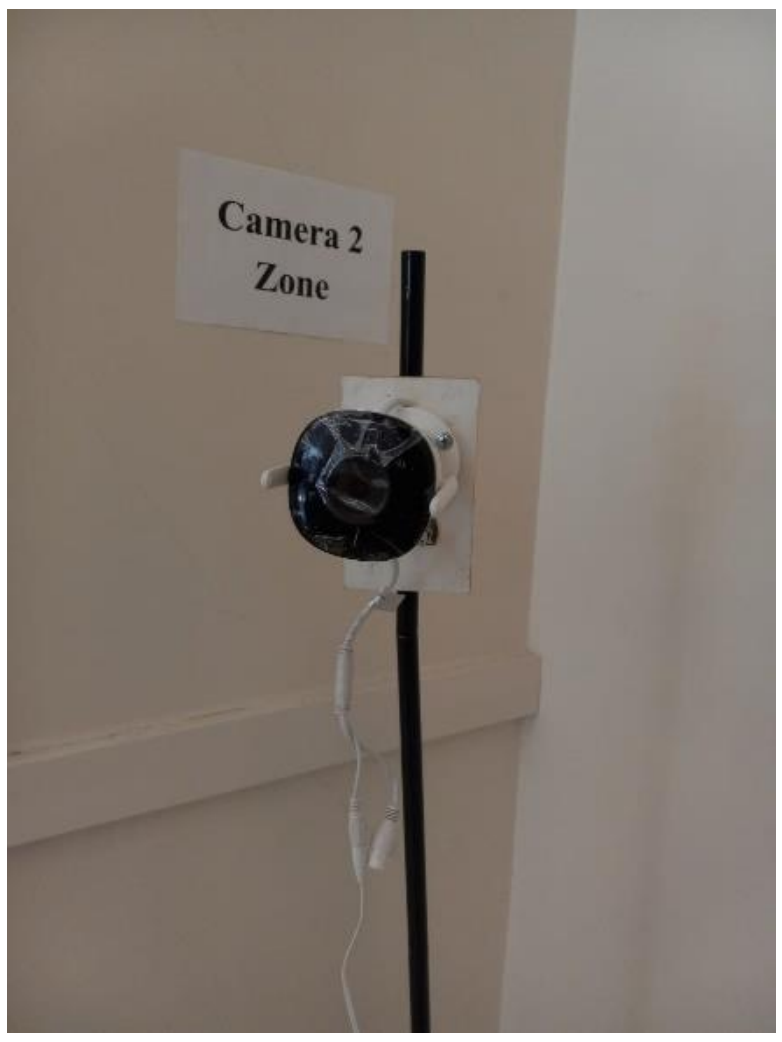

**Figure 7**. Placement of the second Imou Smart Camera

**Table 1**. Costs of the sensors that are necessary for development of the proposed system.

| Sensor | Cost per unit (in USD) | Total Cost (in USD) |
|---|---|---|
| Imou Smart Camera | $29.99 | $59.98 |
| Mbient Labs Sleeve Sensor | $103.99 | $103.99 |
| Proximity Sensors | $24.75 | $99.00 |
| Microsoft SQL Server 11.0 | $0.00 | $0.00 |
| The total cost of all the sensors | | $262.97 |

## 4. Results and Discussions

This section presents the implementation details, preliminary results, and associated discussions to uphold the potential, effectiveness, and relevance of the proposed AAL system. Section 4.1 presents the findings from real-world experiments. This section also compares the functionality of our system with prior works in this field to highlight the novelty of the same. Section 4.2 presents a comparative study with prior works related to fall detection, indoor localization, and assisted living to justify the cost-effective nature of the system.

*4.1. Results and Findings from Real-World Implementation*

The real-world development, implementation, and testing of this system were performed by taking into consideration the safety and protection of human subjects. Therefore, the CITI Training provided by the University of Cincinnati that comprises the CITI Training Curriculum of the Greater Cincinnati Academic and Regional Health Centers [94] was completed by both authors. Thereafter, approval for this study was obtained from the University of Cincinnati's Institutional Review Board (IRB) [95] with IRB Registration #: 00000180, FWA #: 000003152, and the IRB ID for this study was 2019-1026. Figure 8 shows the spatial mapping of two' activity-based zones' in the simulated IoT-based environment. Figures 9-13 show the images captured from the recordings from the Imou smart cameras during one of the participants performing different ADLs in these 'activity-based zones'. The experiment protocol included the participant navigating across different 'activity-based zones' and performing different ADLs, including falls and fall-like motions to simulate real-world ADLs and falls. The performance of the ADLs included different behaviors such as walking, stopping, falling, lying down, getting up from lying, etc., some of which are represented in these figures. These activities were tracked by the Imou Smart Camera's, Mbient Labs Sleeve Sensor, and the Proximity Sensors. The Mbient Labs Sleeve Sensor was mounted on the participant's chest during the entire duration of the experimental trials. The decision to mount an accelerometer on a participant's chest was made as per the findings of Gjoreksi et al. [96]. These findings state that mounting an accelerometer on an individual's chest is the optimal location for accelerometer-based fall detection. The Imou smart cameras were placed at two strategic and calculated locations that helped to track all ADLs performed in all the simulated 'activity-based zones'. The proximity sensors helped to detect the participant's presence or absence in each of these 'activity-based zones' in the context of different behavioral patterns during different ADLs. The software that comprised the Big Data processing and integration module (Figure 1) communicated with all these sensors to capture the associated data, convert it into a common format (.csv files), and develop a database on MS SQL Server with the different data represented as different attributes in the database.

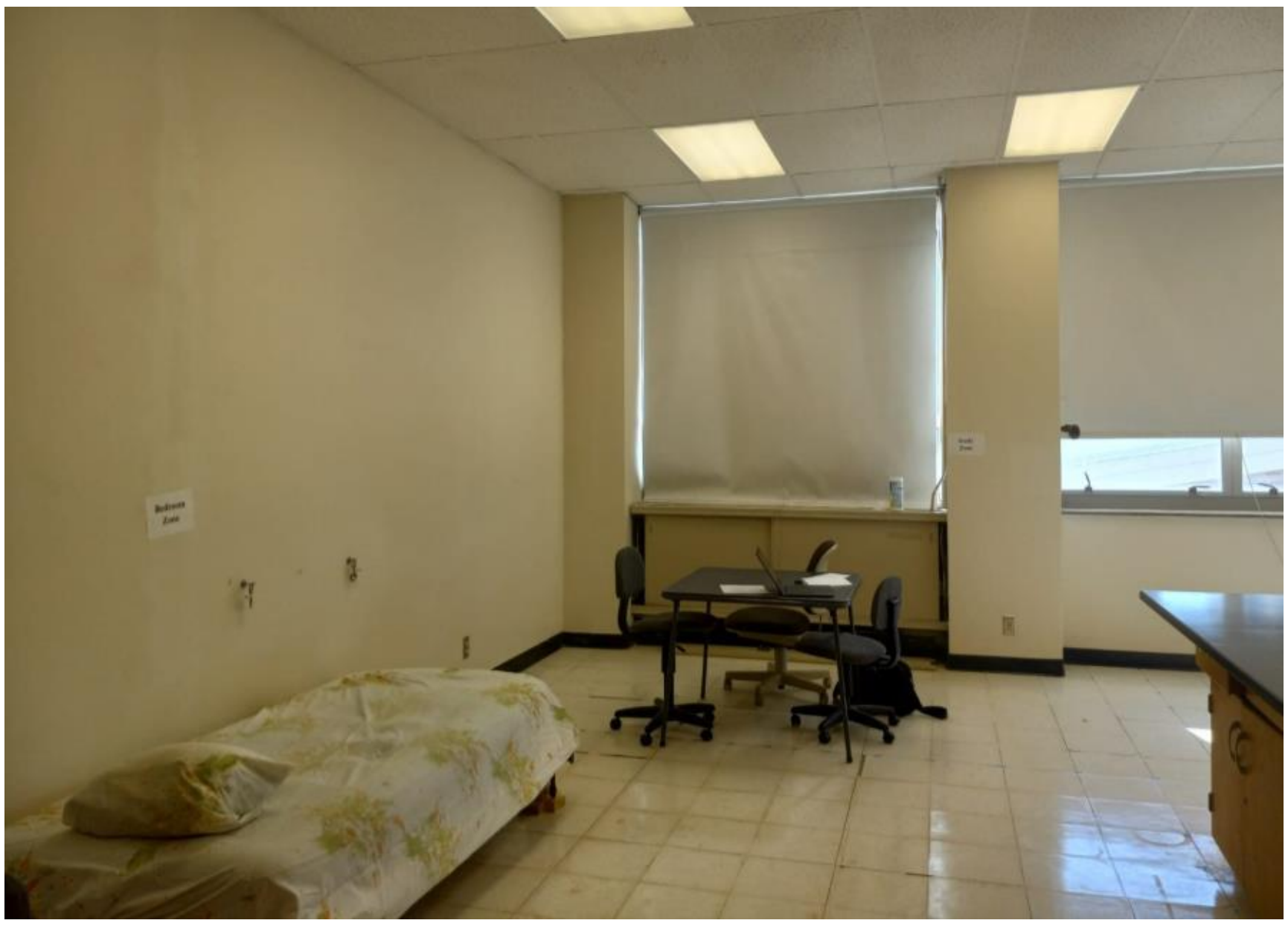

**Figure 8**. Overview of the lab space showing two 'activity-based zones' (bedroom zone and study zone) with multiple contextual parameters

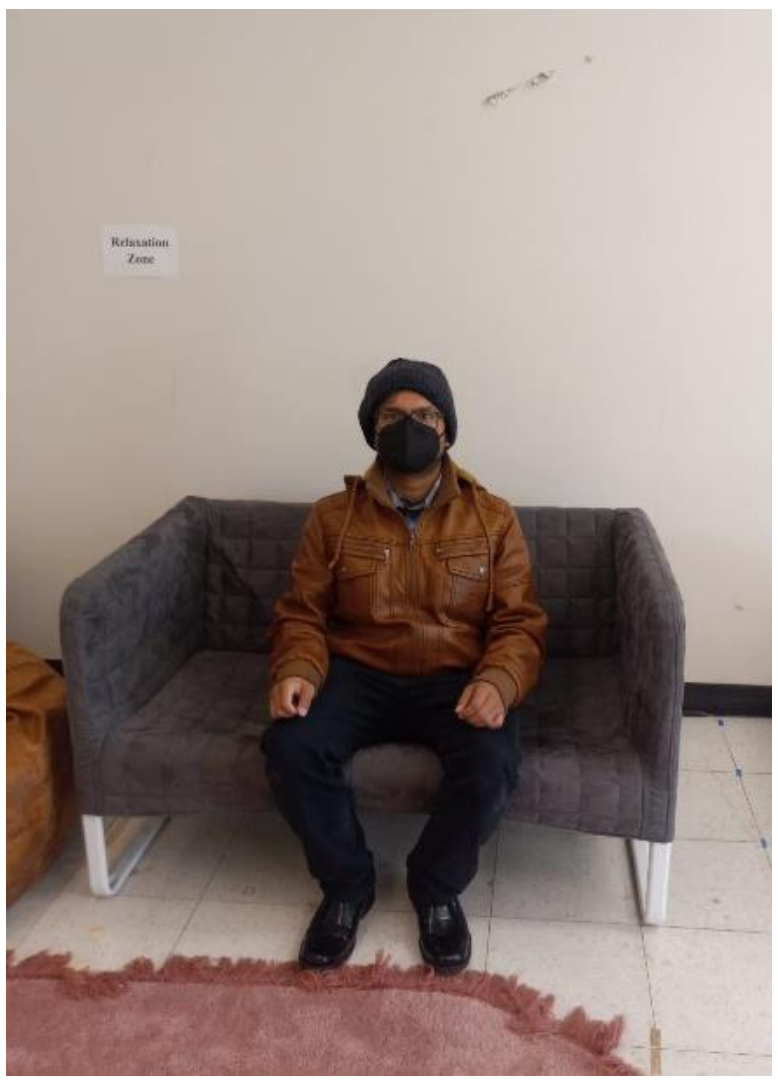

**Figure 9**. A participant sitting on a couch in the relaxation zone

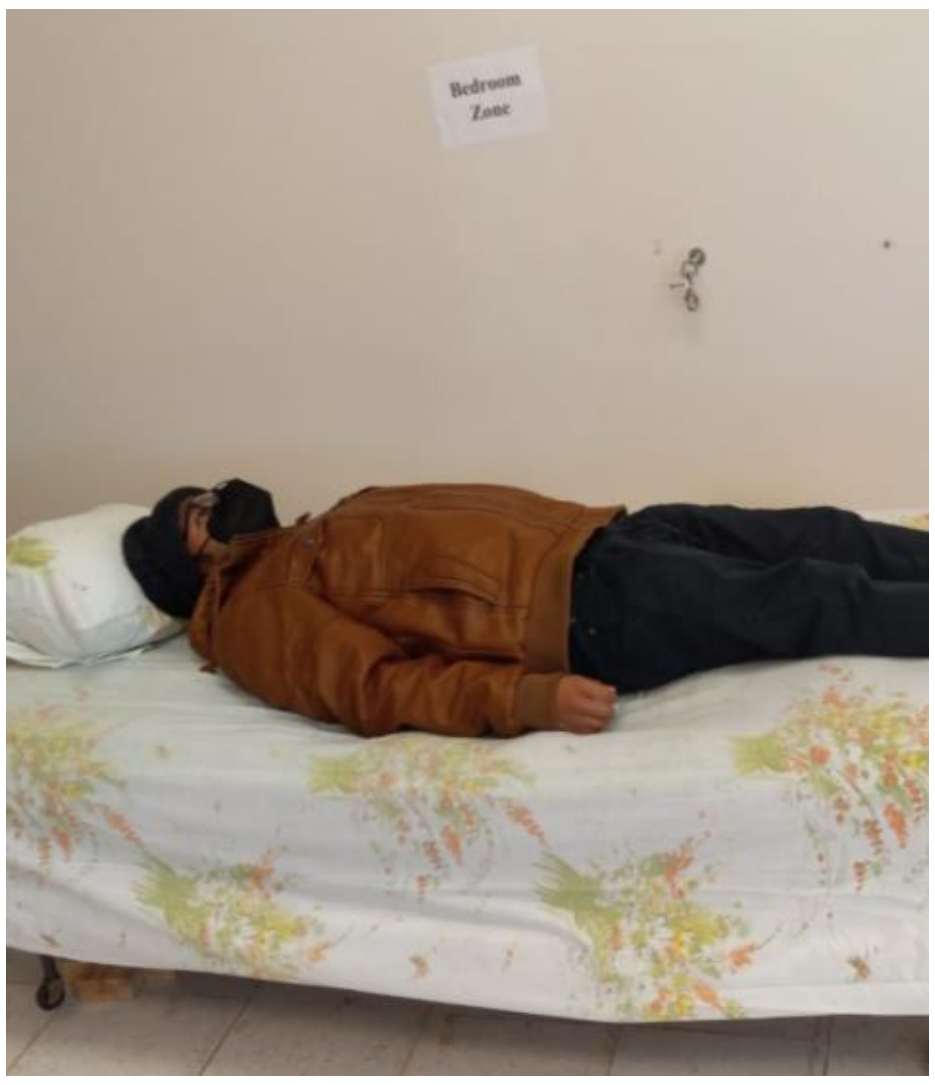

**Figure 10**. A participant lying in the bedroom zone

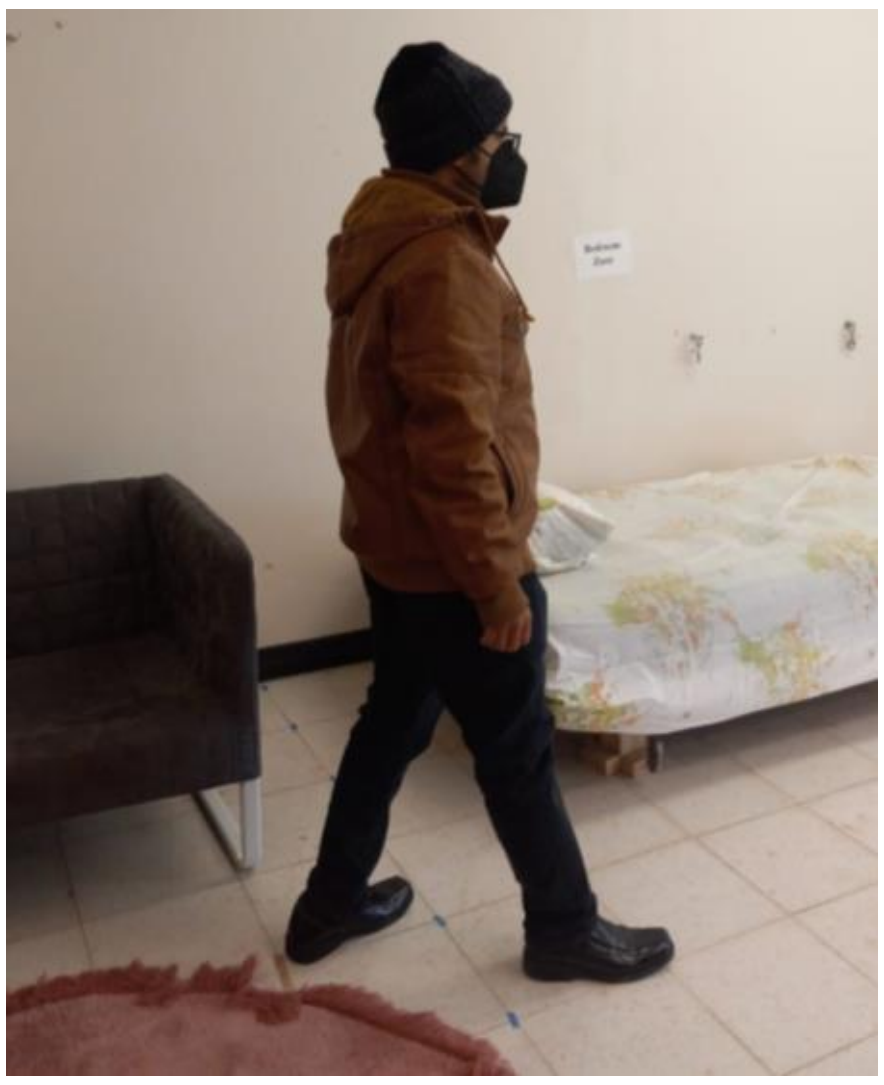

**Figure 11**. A participant walking from the relaxation zone to the bedroom zone

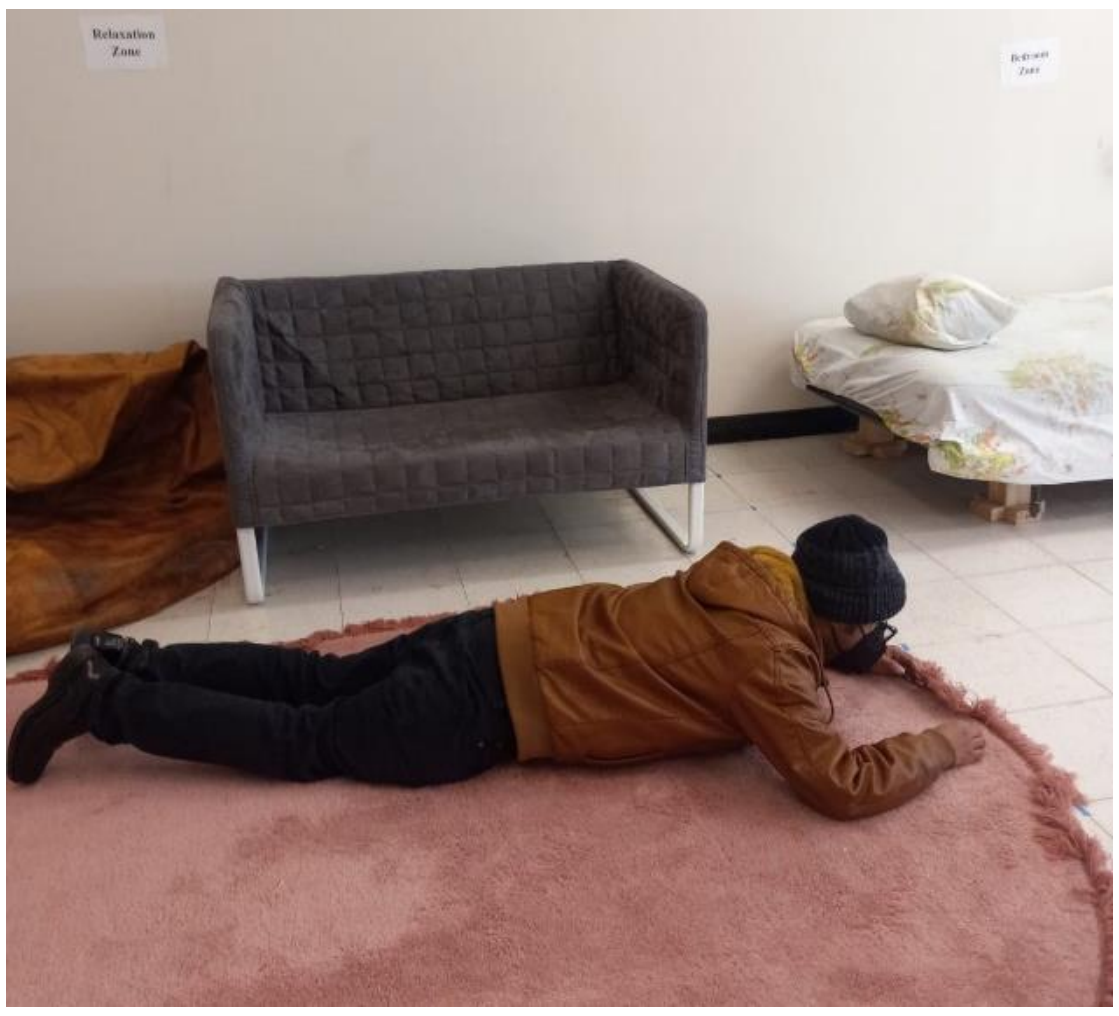

**Figure 12**. A participant falling in the relaxation zone. Here, a forward fall is demonstrated.

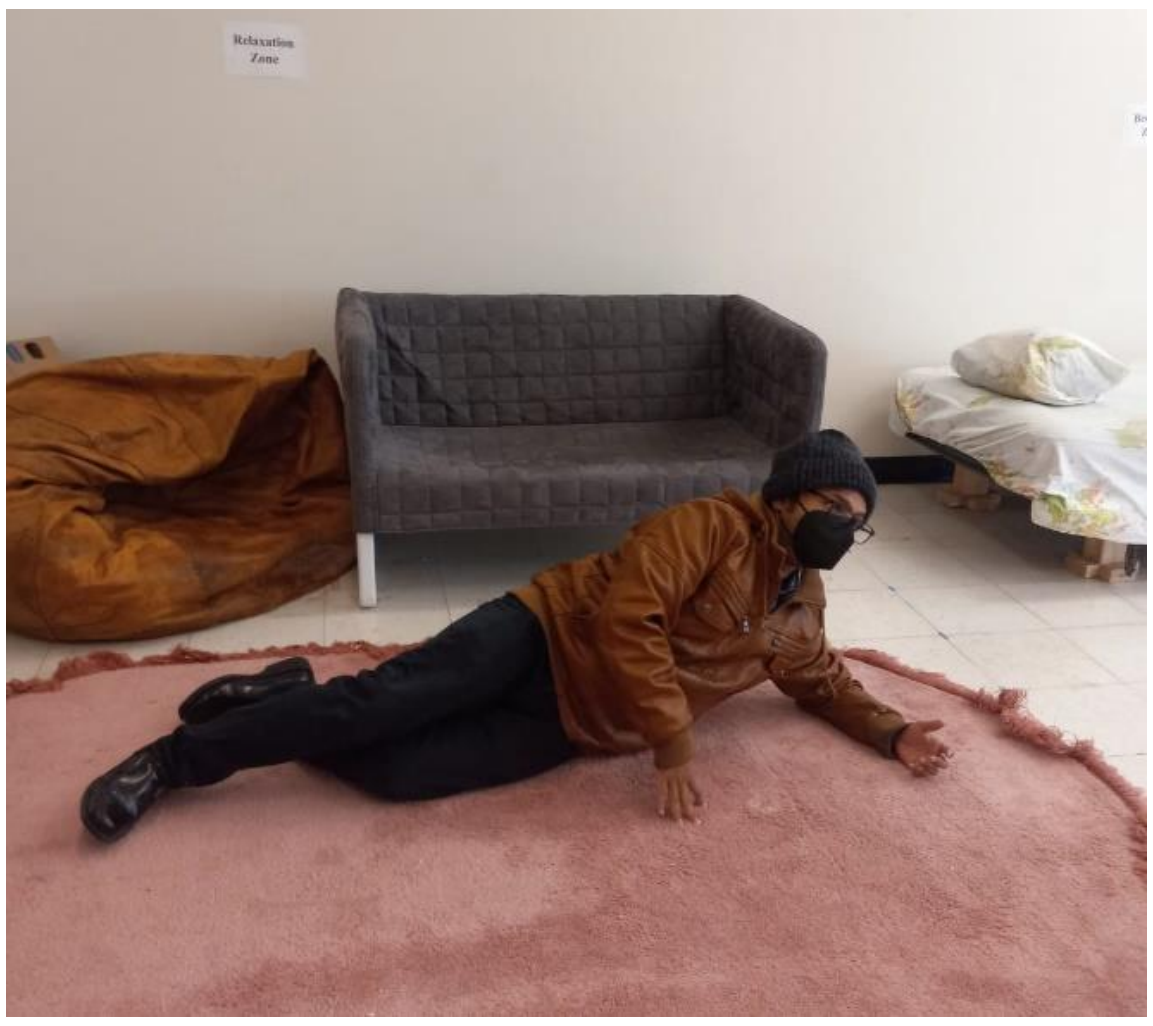

**Figure 13**. A participant falling in the relaxation zone. Here, a sideways fall is demonstrated.

The original plan was to test all the proposed functionalities of this system by including 20 volunteers (ten males and ten females) for user diversity [97]. But no one predicted the widespread of the COVID-19 pandemic [98], the declaration of a national emergency in the United States on account of the same [99], and the subsequent lockdown for several months with more than 517,572,198 cases of COVID-19 with 6,277,510 deaths on a global scale [100]. In the United States alone, at the time of writing this paper, there have been 83,644,452 cases of COVID-19 with 1,024,655 deaths [100]. Therefore, it was not possible to recruit 20 volunteers. As a result, we were able to perform the proposed experiments with only one volunteer. This is one of the limitations of this study, which we plan to address in the near future by recruiting more volunteers for the experiments. While this is a limitation of this study, upon reviewing prior works in this field, we observed that there have been several real-world systems for fall detection, indoor localization, and assisted living, the effectiveness of which was evaluated by one volunteer or participant. Some examples include the works of Godfrey et al. [47], Liu et al. [48], Dinh et al. [49,50], Townsend et al. [51], Dombroski et al. [33], and Ichikari et al. [34] (VDR Track), just to name a few. This is further illustrated in Table 2.

**Table 2**. Prior works in the field of fall detection, indoor location, and assisted living that were tested using the data from only one human subject.

| **Work** | **Number of human subjects** |
|---|---|
| Godfrey et al. [47] | 1 |
| Liu et al. [48] | 1 |
| Dinh et al. [49] | 1 |
| Dinh et al. [50] | 1 |
| Townsend et al. [51] | 1 |
| Cordes et al. [52] | 1 |
| Dombroski et al. [33] | 1 |
| Ichikari et al. [34] (VDR Track) | 1 |
| Lemic et al. [35] | 1 |
| Thakur et al. [this work] | 1 |

Based on Table 2, it can be observed that several prior works in this field evaluated and tested the effectiveness of their respective systems and frameworks by conducting experiments that included only one human subject. So, even though the availability of only one human subject is a limitation of this study, it is consistent with several prior works in this field, and the findings can be considered to uphold the potential, effectiveness, and relevance of the proposed system and its functionality. The experiments with this volunteer were conducted as per the guidelines for reducing the spread of COVID-19, recommended by both the Centers for Disease Control [101] and the University of Cincinnati [102]. These guidelines also involved following the University of Cincinnati's recommendations (at the time of conducting these experiments) to wear a mask in indoor environments, including classrooms and research labs, so the participant shown in Figures 9-13 can be seen wearing a mask. Figure 14 is a screenshot from the MS SQL server where the database was developed that consisted of the integration of the data collected from different sensors during the data collection process.

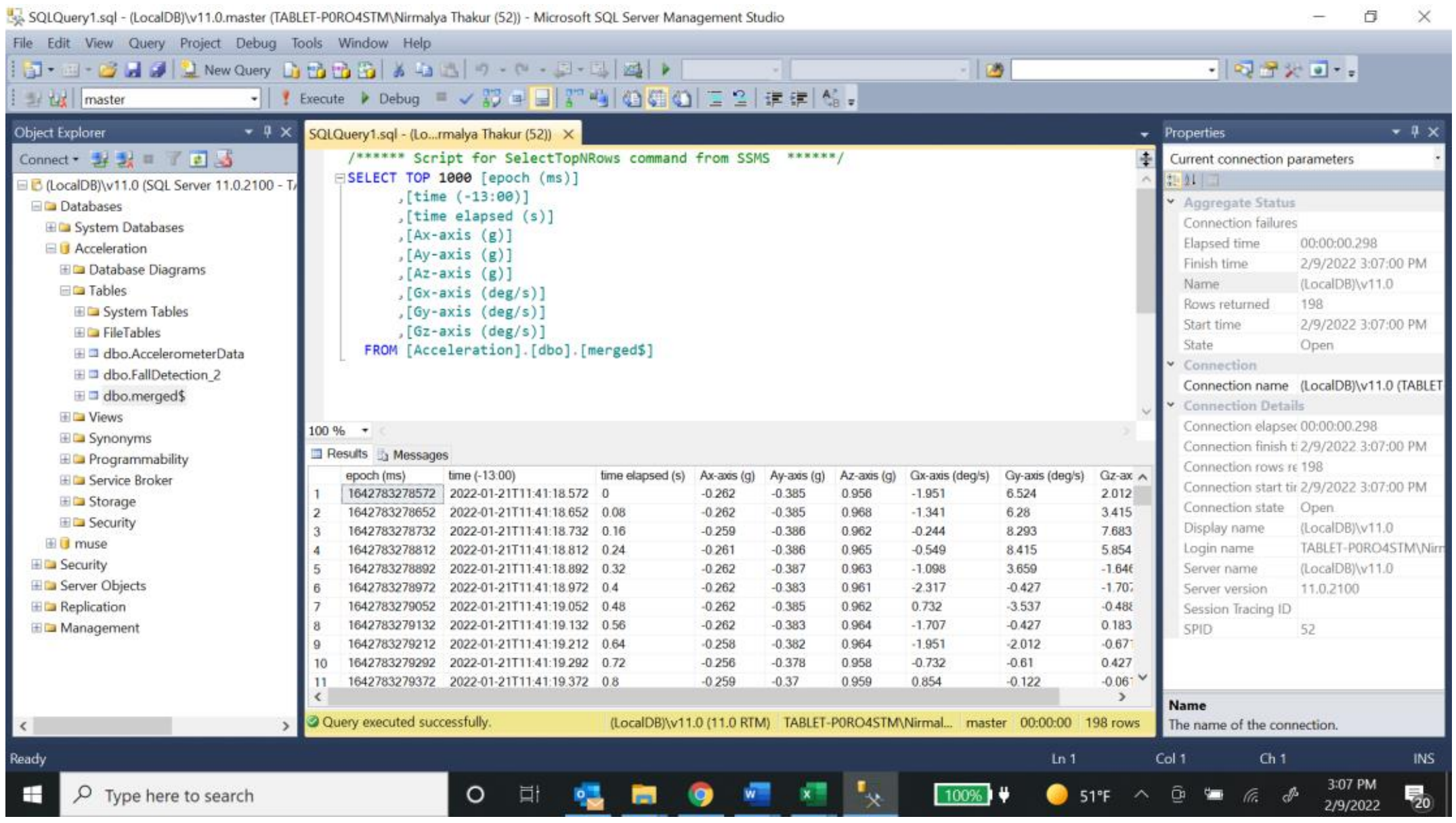

**Figure 14**. Screenshot from the database table in MS SQL Server 2012 version that represents the data collected from different participants during the experimental trials.

In the upper part of Figure 14, the SQL Query is represented that was used to query the developed database. In the lower part of Figure 14, the specific attributes that comprise this dataset are represented. The image shows only the first 11 rows for clarity of representation. The first two attributes represent the time instants. The third attribute represents the time elapsed (in seconds) for the specific behavioral pattern under consideration. The fourth, fifth, and sixth attributes represent the acceleration data recorded along the X, Y, and Z directions for the specific behavioral pattern under consideration. Similarly, the seventh, eighth, and ninth attributes represent the gyroscope data recorded along the X, Y, and Z directions for the specific behavioral pattern under consideration. The data successfully collected from the different sensors during the different ADLs performed in real-time and the successful processing and integration of the same towards the development of this database upholds the relevance and effectiveness of this proposed design methodology. The attributes of this database are the exact same attributes (evidenced from the findings of [84] and [85]) that would be necessary for an AAL system to perform both fall detection and indoor localization during ADLs in a simultaneous manner in real time. Figures 15 and 16 represent the variation of the acceleration data and gyroscope data (in X, Y, and Z directions), respectively, during the different behavioral patterns associated with these ADLs that were performed in different 'activity-based zones' at different time instants. The data analysis to study this variation was performed in RapidMiner [103]. In both these figures, only a few timestamps are shown for clarity of representation.

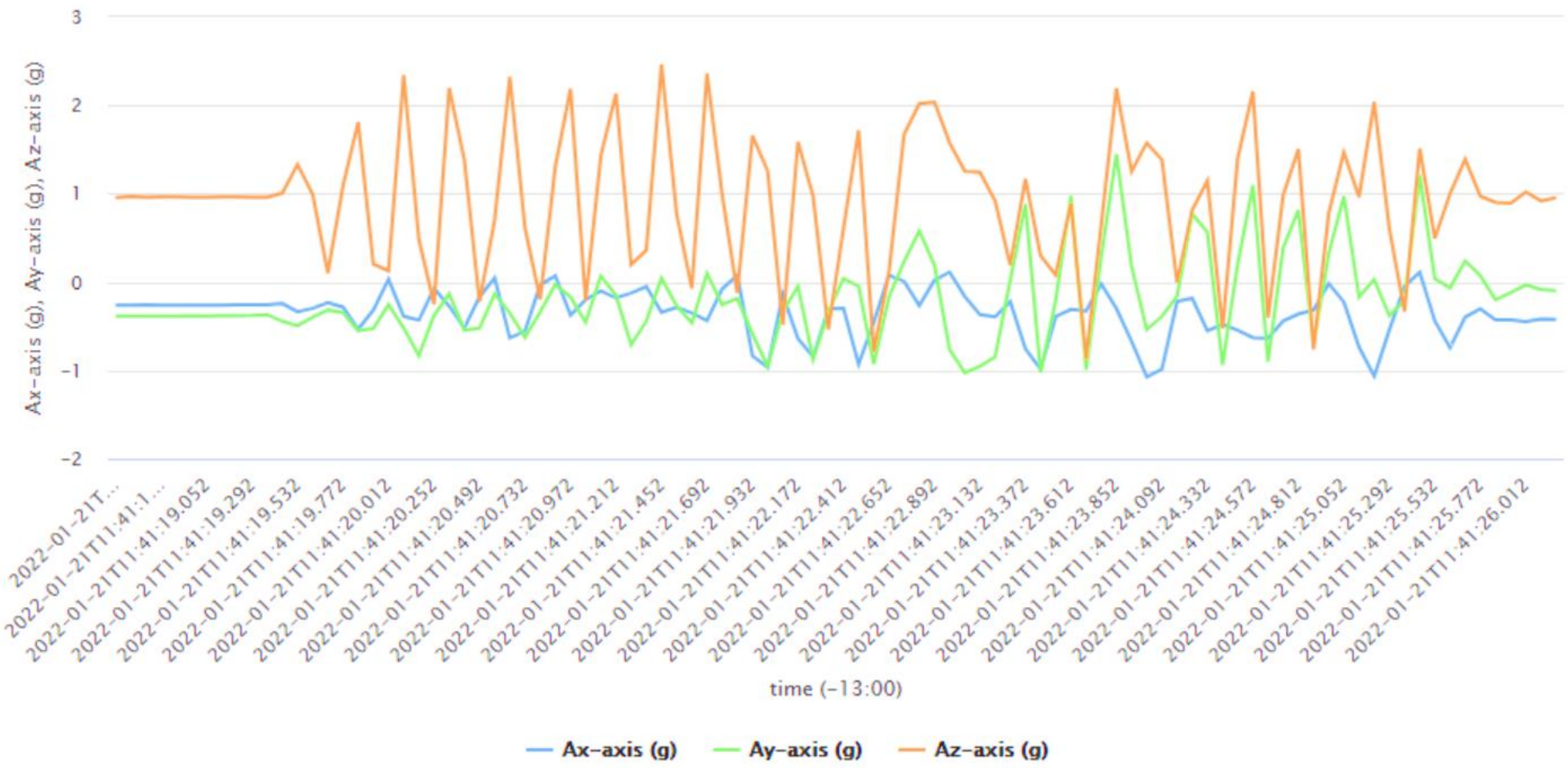

**Figure 15**. Variation of the acceleration data (in X, Y, and Z directions) during the different behavioral patterns associated with different ADLs performed at different time instants.

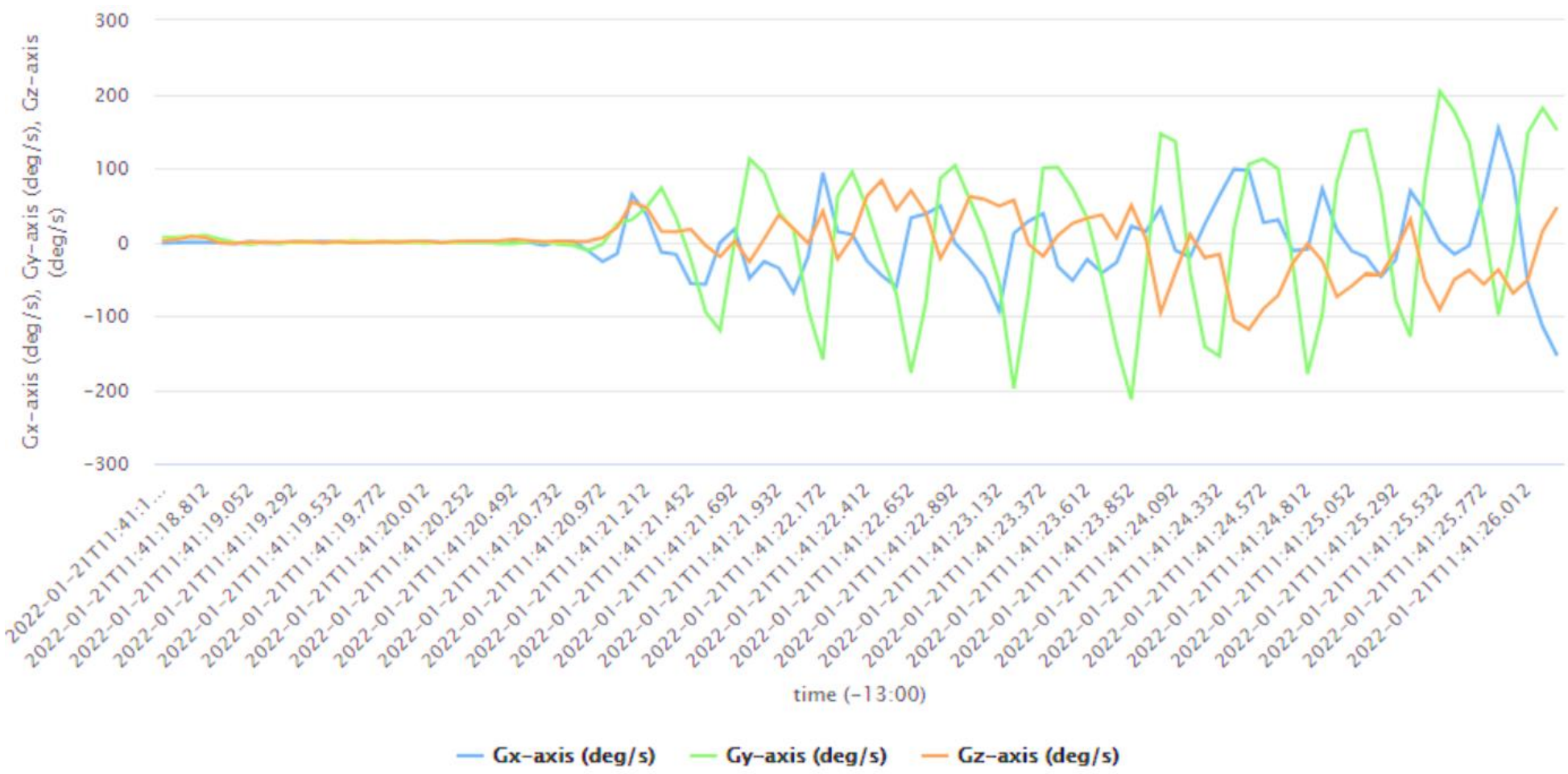

**Figure 16.** Variation of the gyroscope data (in X, Y, and Z directions) during the different behavioral patterns associated with these ADLs performed at different time instants.

The objective of performing these experiments with this volunteer was to demonstrate the working and effectiveness of the proposed system to capture multimodal components of user behaviors during ADLs that are necessary for performing fall detection and indoor localization in a simultaneous manner in the real world. We understand that real-world users of such a system, typically the elderly, are expected to have a range of user diversity characteristics that could be different from one or more of the diversity characteristics of this volunteer who participated in the experiments. Several prior works in the areas of studying human behavior from diverse users [104-107] have shown that while human behavior is affected by user diversity in more than one ways but user diversity has no effect on the working and effectiveness of hardware and software components of a system aimed at collecting multimodal components of human behavior. For instance, a smart watch whose functionalities might have been tested using young volunteers during the development stage will not stop working effectively if its real-world users are elderly people or people with user diversity characteristics that might be different from these young volunteers in one or more ways. Therefore, it can be concluded that this section demonstrates the working and effectiveness of our system in the real world. In Table 3, we compare the functionality of our system with prior works in this field to highlight the novelty of the same.

In Table 3, in terms of effective software design, we refer to the effectiveness (marked by a ✓ sign in the Table) of the underlining software component of the system to receive (either from real-world data collection or from a relevant dataset), integrate, and process the required data that is necessary for the core functionality of the underlining system such as fall detection or indoor localization. In terms of effective hardware design, we refer to the effectiveness (marked by a ✓ sign in the Table) of the hardware components (such as sensors or other equipment) to function in real-time to collect the necessary data (relevant for the core functionality of the system such as fall detection or indoor localization) that can be received by the software component of the system. As can be seen from Table 3, there are several systems that focus on fall detection with an effective software and hardware design. However, the hardware design of these systems does not provide the relevant data to its software components that be integrated and processed by the same to infer the indoor location of the underlining users. For instance, the hardware design of the work in [56] provides the data collected from ultrasonic sensors and field-programmable gate array (FPGA) processors, which the software component of the system can receive, integrate, and process to detect falls, but this data is irrelevant for performing indoor localization. Thus the software design and hardware design of this system and similar works are considered ineffective for indoor localization. Similarly, as can be seen from Table 3, there are several systems that focus on indoor localization with effective software and hardware design. However, the hardware design of these systems does not provide the relevant data to its software components that be integrated and processed by the same to perform fall detection. For instance, the hardware design of the works in [26] and [30] collect the WiFi fingerprint information, which the software component of the system can receive, integrate, and process to perform indoor localization, but this data is irrelevant for performing fall detection. Thus the software design and hardware design of these two systems and similar works are considered ineffective for fall detection. In Table 3, we have also shown how the software design and hardware design of prior works in this field that were evaluated using datasets [58-80,84,85] can be compared with the proposed system. The limitation in such works, including two of our prior works [84,85] (reviewed in Sections 3.1 and 3.2), is that these works present effective software designs with no hardware design. Furthermore, the underlying software designs cannot perform both indoor localization and fall detection simultaneously. The hardware design of our system is able to collect the necessary data (Sections 3.1, 3.2, 3.3, and 4.1) for both fall detection and indoor localization. The software design of our system demonstrates its effectiveness (Section 3.3 and 4.1) in receiving, integrating, and processing the data provided by the hardware component of our system. This upholds the novelty of our system in terms of having an effective software design and hardware design, which can work together and capture multimodal components of the user behavior during ADLs that are necessary for performing fall detection and indoor localization in a simultaneous manner in the real world.

**Table 3**. Comparison of the functionality of our system with prior works in this field in terms of the effectiveness of the software design and hardware design.

| **Focus Area of Work** | **Author Names** | **Fall Detection** | | **Indoor Localization** | |
|---|---|---|---|---|---|
| | | Effective Software Design | Effective Hardware Design | Effective Software Design | Effective Hardware Design |
| Indoor Localization | Varma et al. [23] | | | ✓ | ✓ |
| Indoor Localization | Qin et al. [24] | | | ✓ | |
| Indoor Localization | Musa et al. [25] | | | ✓ | ✓ |
| Indoor Localization | Yim et al. [26] | | | ✓ | ✓ |
| Indoor Localization | Hu et al. [27] | | | ✓ | ✓ |
| Indoor Localization | Poulose et al. [28] | | | ✓ | ✓ |
| Indoor Localization | Barsocchi et al. [29] | | | ✓ | ✓ |
| Indoor Localization | Kothari et al. [30] | | | ✓ | ✓ |
| Indoor Localization | Wu et al. [31] | | | ✓ | ✓ |
| Indoor Localization | Gu et al. [32] | | | ✓ | ✓ |
| Fall Detection | Rafferty et al. [43] | ✓ | ✓ | | |
| Fall Detection | Ozcan et al. [44] | ✓ | ✓ | | |
| Fall Detection | Khan et al. [45] | ✓ | ✓ | | |
| Fall Detection | Cahoolessur et al. [46] | ✓ | ✓ | | |
| Fall Detection | Godfrey et al. [47] | ✓ | ✓ | | |
| Fall Detection | Liu et al. [48] | ✓ | ✓ | | |
| Fall Detection | Dinh et al. [49,50] | ✓ | ✓ | | |
| Fall Detection | Townsend et al. [51] | ✓ | ✓ | | |
| Fall Detection | Hsu et al. [53] | ✓ | ✓ | | |
| Fall Detection | Yun et al. [54] | ✓ | ✓ | | |
| Fall Detection | Nguyen et al. [55] | ✓ | ✓ | | |
| Fall Detection | Huang et al. [56] | ✓ | ✓ | | |
| Indoor Localization | Song et al. [58] | | | ✓ | |
| Indoor Localization | Kim et al. [59] | | | ✓ | |
| Indoor Localization | Jang et al. [60] | | | ✓ | |
| Indoor Localization | Wang et al. [61] | | | ✓ | |
| Indoor Localization | Qin et al. [62] | | | ✓ | |
| Indoor Localization | Wietrzykowski et al. [63] | | | ✓ | |
| Indoor Localization | Panja et al. [64] | | | ✓ | |
| Indoor Localization | Yin et al. [65] | | | ✓ | |
| Indoor Localization | Patil et al. [66] | | | ✓ | |
| Indoor Localization | Gan et al. [67] | | | ✓ | |
| Indoor Localization | Hoang et al. [68] | | | ✓ | |
| Indoor Localization | Seçkin et al. [69] | | | ✓ | |
| Fall Detection | Galvão et al. [70] | ✓ | | | |
| Fall Detection | Sase et al. [71], | ✓ | | | |
| Fall Detection | Li et al. [72], | ✓ | | | |
| Fall Detection | Theodoridis et al. [73], | ✓ | | | |
| Fall Detection | Abobakr et al. [74], | ✓ | | | |
| Fall Detection | Abdo et al. [75], | ✓ | | | |
| Fall Detection | Sowmyayani et al. [76], | ✓ | | | |
| Fall Detection | Kalita et al. [77], | ✓ | | | |
| Fall Detection | Soni et al. [78], | ✓ | | | |
| Fall Detection | Serpa et al. [79], | ✓ | | | |
| Fall Detection | Lin et al. [80] | ✓ | | | |
| Fall Detection | Thakur et al. [84] | ✓ | | | |
| Indoor Localization | Thakur et al. [85] | | | ✓ | |
| Indoor Localization and Fall Detection | Thakur et al. [this work] | ✓ | ✓ | ✓ | ✓ |

*4.2. Comparative Study to Uphold the Cost-Effectiveness of the System*

As discussed in Section 2, there have been a few prior works related to the development of fall detection, indoor localization, and assisted living technologies that focused on real-world development and implementation. However, one of the major limitations of these works is the high cost (primarily cost of the underlining equipment) of the underlining systems, which have posed a major challenge to the wide-scale deployment of the same across multiple IoT-based environments. In addition to proposing a simplistic design for our system, we also aimed to address the research challenge of developing a cost-effective system in this study. There can be multiple ways by which the cost of a system may be defined for different use-case scenarios. These could include the cost of equipment, cost of installation, cost of maintenance, the salary of research personnel, cost of deployment, computational costs, etc. The cost of a system can also be defined as a combination of two or more of these components. In this section, we compare the cost of our system with the cost of prior works in this field to uphold its cost-effective nature. All these types of costs, as mentioned above, could be interesting components to computer and/or compare. However, most of the prior works in these fields reported only the cost of equipment, so only the cost of necessary equipment was used as the grounds for comparison in this comparative study. Furthermore, comparing the cost of the associated equipment to comment on the cost-effectiveness of the underlying system is an approach that has been followed by researchers in the broad domain of IoT as can be seen from recent works [108-111] in this field. As shown in Table 1, the cost of equipment for the development of this AAL system was USD 262.97. Table 4 shows the comparison of the costs of our system with prior works in this field.

**Table 4**. Comparison of the costs of the proposed system with the cost of the systems proposed in prior works in this field. Here, by costs, we refer to only the cost of equipment.

| **Work** | **Costs (in USD)** |
|---|---|
| Muffert al. [37] | >$10,000 |
| Kohoutek et al. [36] | $9,000 |
| Popescu et al. [40] | $1,500.00 |
| Yun et al. [54] | $1372.49* |
| Tilch et al. [38] | $ 1,055.98* |
| Habbecke et al. [39] | $1,055.98* |
| Hsu et al. [53] | $950.21* |
| Liu et al. [41] | $844.61* |
| Huang et al. [56] | $750.00 |
| Dasios et al. [57] | $581.00 |
| Braun et al. [42] | $460.00 |
| Nguyen et al. [55] | $422.36* |
| Thakur et al. [this work] | $262.97 |

In Table 4, * denotes the USD equivalent of Euros (as per the conversion rate mentioned on https://www.xe.com/ on May 9, 2022). This conversion to USD was performed for consistency in this comparative study in terms of comparing the costs of all the systems in USD, as the costs of some of these systems [54,38,39,53,41, and 55] were reported in Euros in the associated publications. As can be seen from Table 4, the development and implementation of this proposed system involves the least cost as compared to all prior works in the fields of fall detection, indoor localization, and assisted living, which involved real-world development. This upholds the cost-effective nature of the proposed system and also illustrates the potential of the same for wide-scale real-world implementation in multiple smart homes for AAL of the elderly.

## 5. Conclusions and Future Work

As the population ages, modern society is facing a wide range of difficulties stemming from numerous conditions and needs associated with the elderly. Over the last few years, the aging populations throughout the globe have had to contend with a decrease of caregivers to care for them, which has created a variety of difficulties related to independence in carrying out ADLs, which are crucial for one's sustenance. Falls are highly common in the elderly on account of the various bodily limitations, challenges, declining abilities, and decreasing skills that they face with increasing age. Falls can have a variety of negative effects on the health, well-being, and quality of life of the elderly, including restricting their capability to perform or complete ADLs. Falls can even lead to permanent disabilities or death in the absence of timely care and services. To detect these dynamic and diversified needs of the elderly that usually arise in the context of their living environments during ADLs, tracking, and analysis of the spatial and contextual data associated with these activities, or in other words, indoor localization, becomes very crucial.

Despite several works in the fields of fall detection and indoor localization, there exist multiple challenges centered around (1) lack of fall detection systems to perform indoor localization and vice versa; (2) high cost associated with the real-time development and deployment of the underlining systems; (3) complicated design paradigms of the associated systems; (4) testing of the proposed systems on datasets and lack of guidelines for real-world development of the same; and (5) the systems requiring the development of new devices or gadgets which in addition to being costly, pose a challenge for wide-scale development and deployment in multiple real-world settings. To address these challenges, this work proposes a cost-effective and simplistic design paradigm for an Ambient Assisted Living system that can capture multimodal components of the user behavior during ADLs that are necessary for performing fall detection and indoor localization in the real world. The total cost (in terms of cost of equipment) for the development and implementation of this system is $262.97, which is the least compared to the cost of prior works in the fields of fall detection, indoor localization, and assisted living, which involved real-world development. This upholds the cost-effective nature of the proposed system. The paper also presents a comprehensive comparative study with prior works in this field to highlight the novelty of the proposed AAL system in terms of effective software design and hardware design. Future work would involve the recruitment of more volunteers for the experiments and adding an additional module to this system that would incorporate real-world decision-making in terms of detecting a fall and the associated indoor location of the user during ADLs.

**Author Contributions:** Conceptualization, NT; methodology, NT; software, NT; validation, NT; formal analysis, NT; investigation, NT; resources, NT; data curation, NT; visualization, NT; data analysis and results, NT; writing—original draft preparation, NT; writing—review and editing, NT; supervision, CYH; project administration, CYH; funding acquisition, Not Applicable. All authors have read and agreed to the published version of the manuscript.

**Funding:** This research received no external funding

**Institutional Review Board Statement:** The study, with IRB ID 2019-1026, was approved by the University of Cincinnati's Institutional Review Board (IRB) with IRB Registration#: 00000180 and FWA #: 000003152.

**Informed Consent Statement:** All subjects gave their informed consent for inclusion before they participated in the study.

**Data Availability Statement:** The data presented in this study are available on request from the corresponding author.

**Conflicts of Interest:** The authors declare no conflict of interest.